\documentstyle[12pt]{article}
\begin{document}
\setlength{\baselineskip}{24pt}
\title{ Simultaneous Projectile-Target Excitation in Heavy Ion Collisions}
\author{C.J. Benesh and J.L. Friar,\\
Theoretical Division, MSB283,\\
Los Alamos National Laboratory, \\
Los Alamos, NM 87545}

\maketitle

\begin{abstract}
	We calculate the lowest-order contribution to the cross section
for simultaneous excitation of projectile and target nuclei in relativistic
heavy ion collisions. This process is, to leading order, non-classical and
adds incoherently to the well-studied semi-classical Weizs\"acker-Williams
cross section. While the leading contribution to the cross section is down
by only $1/Z_P$ from the semiclassical process, and consequently of potential
importance for understanding data from light projectiles, we find that phase
space considerations render the cross section utterly negligible.
\end{abstract}
\vfill\eject

	Increasingly, heavy ion beams are being exploited as a tool for
measuring electromagnetic cross sections for use in studies
of nuclear structure\cite{1} and astrophysics\cite{2}. Central to this effort
is the use of the semi-classical Weizs\"acker-Williams method\cite{3}
to relate
the cross section measured in heavy ion collisions to those obtained with
real photons. In this report, we continue a program of examining the
the corrections to the semiclassical approach\cite{4} by calculating the
leading-order contribution to the cross section for simultaneous
excitation of both the target and projectile nucleus.

	Before embarking on the detailed calculation, it is useful to
perform a simplistic analysis of the processes shown in Fig. 1. In Figs. 1a
and b, the leading order Feynman diagrams for target excitation with and
without an accompanying excitation of the projectile are shown.
In the latter case, the projectile form factor is
normalized by the charge of the projectile,
$Z_p$, while in 1a, the transition form factors for the projectile can be
thought of as
being normalized by an appropriately chosen energy-weighted sum rule($\propto
Z_p^{1/2}$). Since the final
states in Figs. 1a and b are distinct, the simultaneous excitation amplitude
adds incoherently to that for single excitation, leading to the
expectation that the semi-classical cross section for single excitation
will lie below the measured cross sections. (This effect is seen in
single neutron removal data taken with low-$Z$ projectiles\cite{1},
but is most likely due to the difficulties inherent in
estimating the strong-interaction contribution to the measured cross
sections\cite{5}.) Since
the cross section for simultaneous excitation is down from that for single
excitation by a full power of $Z_P$, the simultaneous excitation process
is not relevant for high-$Z$ projectiles, but, at least at first glance,
may play a nonnegligible role for low-$Z$ projectiles.

	In Fig. 1b, the projectile, of mass $M_P$ and
momentum $P_i^\mu=(E_i,\vec P_i)$, scatters elastically from the target by
exchanging
a virtual photon of momentum $q^\mu$. In the process, the target, of mass
$M_T$, is excited from its ground state to an excited state of mass
$M_T+\omega_T$. In Fig. 1a, the picture is essentially the same, except
that the projectile is excited to a state of mass $M_P+\omega_P$.
Kinematically, this requires that
\begin{equation}
q^2-2P_i\cdot q=2M_P\omega_p +\omega_P^2.
\end{equation}
For nuclear transitions, the momentum transfers and excitation energies
are negligible compared to the masses, and we obtain
\begin{equation}
-\omega_P=q^0_{A}=\gamma(q^0_{L}-\vec\beta\cdot\vec q_L),
\end{equation}
where the subscript $L(A)$ denotes the momentum transfer evaluated in
the target(projectile) rest frame, $\vec\beta=\vec P_i/E_i$, and
$\gamma =1/\sqrt{1-\vec\beta^2}$. Similarly, for the target we obtain
\begin{equation}
\omega_T=q^0_{L}=\gamma(q^0_{A}+\vec\beta\cdot\vec q_A).
\end{equation}
Thus, the minimum three- and four-momentum transfers are given by
\begin{eqnarray}
\vert\vec q^{min}_L\vert &=&q_L^\parallel =(\gamma\omega_T
+\omega_P)/\beta\gamma, \\
\vert\vec q^{min}_A\vert &=&q_A^\parallel =(\gamma\omega_P
+\omega_T)/\beta\gamma, \\
q^2_{min} &=& -(\omega_T^2+\omega_P^2+2\gamma\omega_T\omega_P)/\beta^2\gamma^2,
\end{eqnarray}
where $q^\parallel$ refers to the component of $\vec q$ along $\beta$.
In contrast to the case of single excitation, where $q^2_{min}$ varies like
$\gamma^{-2}$, the minimum value of $q^2$ drops only like $\gamma^{-1}$, so
that the cross section for simultaneous excitation is less sensitive to
the pole in the photon propagator. Consequently, the Coulomb
contribution to the cross section will be more important than in the single
excitation case.

	 Evaluating the contribution to the spin averaged/summed cross section
from Fig. 1a, we obtain
\begin{eqnarray}
d\sigma_{SE}=
{e^4\over 4((P_i\cdot K_i)^2-M_P^2M_T^2)^{1/2}} {d^3\tilde P_f
d^3\tilde K_f\over q^4}(2\pi)^4\delta(P_f+K_f-P_i-K_i)& &\nonumber\\
\times\sum_{M_i^P,M_f^P} {\langle P_f M_f^P|J_\mu(0)|P_i M_i^P\rangle
\langle P_i M_i^P|J_\nu(0)|P_f M_f^P\rangle\over  (2J_i^P+1)}\nonumber\\
\times\sum_{M_i^T,M_f^T}{\langle K_f M_f^T|J_\nu(0)
|K_i M_i^T\rangle\langle K_i M_i^T|J^\mu(0)|K_f M_f^T\rangle\over (2J_i^T+1)}
& &\nonumber \ ,\\
\end{eqnarray}
where $J_{i}^{T(P)}$ is the initial state target(projectile) spin and
$M_{i(f)}^{T(P)}$ is the third component of the initial(final) state spin.
Following reference 4, we replace the spin sums by Lorentz covariant
structure tensors and rewrite the cross section in terms of matrix elements
defined in the projectile and target rest frames. The result is
\begin{eqnarray}
\sigma_{SE} &=& {e^4\over 4\beta\gamma M_PM_T}\int {d^4q\over (2\pi)^4}
{1\over q^4}\int d^3\tilde P_f d^3\tilde K_f (2\pi)^4\delta^4(P_f-P_i+q)
\nonumber\\
& &\times(2\pi)^4\delta^4(K_f-K_i-q)\Bigg{[}\overline{(\rho\rho)}_P
\overline {(\rho\rho)}_T\left ( {9\over 4}{(\gamma q^2
+\omega_P\omega_T)^2q^4\over
\vec q_A^4\vec q_L^4}-{3\over 4}{q^4\over \vec q_L^2\vec q_A^2}\right )
\nonumber\\
& & \qquad\qquad\qquad\qquad + \overline{(\rho\rho)}_P
\overline{(\rho\rho-\vec J\cdot\vec J)}_T\left (
-{1\over 4}{q^2\over \vec q_A^2}+{3\over 4}{(\gamma q^2 +\omega_p\omega_T)^2
\over \vec q_A^2\vec q_L^2}\right )\nonumber\\
& & \qquad\qquad\qquad\qquad + \overline{(\rho\rho)}_T
\overline{(\rho\rho-\vec J\cdot\vec J)}_P\left (
-{1\over 4}{q^2\over\vec q_L^2}+{3\over 4}{(\gamma q^2 +\omega_p\omega_T)^2
\over \vec q_L^2\vec q_A^2}\right )\nonumber\\
& & \qquad\qquad\qquad +\overline{(\rho\rho-\vec J\cdot\vec J)}_P
\overline{(\rho\rho-\vec J\cdot\vec J)}_T\left
({1\over 4}+{1\over 4}{(\gamma q^2 +\omega_p\omega_T)^2
\over \vec q_L^2\vec q_A^2}\right )\Bigg{]},\nonumber \\
\end{eqnarray}
where $\overline{\rho (\vec J)\rho(\vec J)}_{P(T)}$ indicates the spin-averaged
matrix element of projectile(target)
transition charge(current) density evaluated in the projectile(target)
rest frame.

	 Assuming a model for the transition densities, the remaining
integrations may be carried out numerically or, alternatively, a
useful estimate of the cross section can be obtained in a model-independent
manner by considering the limit of large $\gamma$ and low transition
energies for both the target and projectile. Reexpressing the spatial
delta functions by an integral over complex exponentials and using
translational invariance, the integrals over the components of $\vec q$
orthogonal to $\vec\beta$ can be done analytically, yielding
\begin{eqnarray}
\sigma_{SE} &=& {e^4\over 8\pi M_PM_T}\int \tilde{d^3 P_f}\tilde{d^3 K_f}
\int d^3x d^3y e^{i\,(q_A^\parallel x^\parallel - q_L^\parallel y^\parallel)}
\nonumber\\
& &\times\left [\overline{\rho(\vec x)\rho(0)}_P
\overline{\rho(\vec y)\rho(0)}_T F_{\rho\rho}(\vec z_\perp,q_L^\parallel,
q_A^\parallel,\delta\omega^2)\right .\nonumber\\
& &+ \overline{\rho(\vec x)\rho(0)}_P
\overline{(\rho(\vec y)\rho(0)-\vec J(\vec y)\cdot\vec J(0))}_T
F_{\rho\gamma}(\vec z_\perp,q_A^\parallel,
q_L^\parallel,\sqrt{-q^2_{min}},\delta\omega^2)\nonumber\\
& &+ \overline{\rho(\vec y)\rho(0)}_T
\overline{(\rho(\vec x)\rho(0)-\vec J(\vec x)\cdot\vec J(0))}_P
F_{\rho\gamma}(\vec z_\perp,q_L^\parallel,
q_A^\parallel,\sqrt{-q^2_{min}},-\delta\omega^2)\nonumber\\
& &+ \overline{(\rho(\vec x)\rho(0)-\vec J(\vec x)\cdot\vec J(0))}_P
 \overline{(\rho(\vec y)\rho(0)-\vec J(\vec y)\cdot\vec J(0))}_T
\nonumber\\ & & \times F_{\gamma\gamma}(\vec z_\perp,q_A^\parallel,
q_L^\parallel,\sqrt{-q^2_{min}},\delta\omega^2)\left .\right ],
\end{eqnarray}
where $\vec x,\vec y$ are the arguments of the matrix elements appearing in
the spin averages, $\vec z_\perp$ is the component of $\vec x-\vec y$
orthogonal to $\vec\beta$, $x^\parallel$ and $y^\parallel$ are the components
of
$\vec x$ and $\vec y$ parallel to $\vec \beta$, and
\begin{eqnarray}
F_{\rho\rho}(\vec z_\perp,q_A^\parallel,
q_L^\parallel,\delta\omega^2)\qquad\qquad &=& {9\over 8}\left [
{\omega_P^2\over (\delta\omega^2)^2} q_A^\parallel z_\perp K_1( q_A^\parallel
 z_\perp)+ {\omega_T^2\over (\delta\omega^2)^2} q_L^\parallel z_\perp
K_1( q_L^\parallel z_\perp)\right ]\qquad\qquad\qquad\qquad\nonumber\\
& &-{3(\omega_T^4+\omega_P^4+(6\gamma^2+4)\omega_P^2\omega_T^2+
6\gamma\omega_T\omega_P(\omega_T^2+\omega_P^2))\over 4(\delta\omega^2)^3\beta^2
\gamma^2}\nonumber\\
& &\times\left [K_0(q_A^\parallel z_\perp)-K_0(q_L^\parallel z_\perp)\right ],
\nonumber\\
F_{\rho\gamma}(\vec z_\perp,q_A^\parallel,q_L^\parallel,\sqrt{-q^2_{min}},
\delta\omega^2)&=&{3 q_A^\parallel z_\perp K_1( q_A^\parallel
z_\perp)\over 8\delta\omega^2}\nonumber\\
& &+{1\over 2\omega_P^2\beta^2\gamma^2}
\left [K_0(q_A^\parallel z_\perp)-K_0(q_L^\parallel z_\perp)\right ]
\nonumber\\
& &+{3{q_L^\parallel}^2\over 4(\delta\omega^2)^2}\left [K_0(\sqrt{-q_{min}^2}
z_\perp)-K_0(q_A^\parallel z_\perp)\right ],\nonumber\\
F_{\gamma\gamma}(\vec z_\perp,q_A^\parallel,q_L^\parallel,\sqrt{-q^2_{min}},
\delta\omega^2) &=& {1\over 4}{\sqrt{-q_{min}^2}z_\perp K_1(\sqrt{-q_{min}^2}
z_\perp)
\over
-q_{min}^2\beta^2\gamma^2}+{q_{min}^2K_0(\sqrt{-q_{min}^2}z_\perp)
\over 4\omega_T^2
\omega_P^2}\nonumber\\
& &+{{q_A^\parallel}^2K_0(q_A^\parallel z_\perp)\over 4\delta\omega^2
\omega_P^2}-{{q_L^\parallel}^2K_0(q_L^\parallel z_\perp)\over 4\delta\omega^2
\omega_T^2},\\
\end{eqnarray}
where $K_0,K_1$ are modified Bessel functions, and $\delta\omega^2=\omega_T^2
-\omega_P^2$. The strategy at this point
is to expand the Bessel functions around zero frequency, assuming that the
logarithms of $\vert \vec z_\perp\vert$, $\vert\vec x\vert$, and
$\vert\vec y\vert$ vary slowly enough to be treated as constants. For
dipole-dipole excitations, we obtain, after much algebra
\begin{equation}
 \sigma_{E1E1}={3\over 8\pi^3}\int d\omega_P d\omega_T
{\sigma_{P\gamma}^{E1}(\omega_P)\sigma_{T\gamma}^{E1}(\omega_T)\over\omega_P
\omega_T}[(\omega_P^2+\omega_T^2)\xi_1-\delta\omega^2\xi_2]+{\cal O}(1/\gamma),
\end{equation}
where $\xi_1$ and $\xi_2$ are averages of logarithms of projectile/target
coordinates over transition densities and are both of order one numerically.
Remarkably, all of the dependence on logarithms of the transition frequencies
cancels out of expression 12, so that $\xi_1$ and $\xi_2$ have no explicit
dependence on the transition frequencies. In addition, we note that the
photon-pole terms are
all of order $1/\gamma$, so that the high-$\gamma$ limit of the cross section
is dominated by the off-shell response functions of the target and projectile.

	Assuming that the dipole cross sections are sharply peaked at
the giant resonance energy, and that $\xi_1\approx 1$, we estimate the
simultaneous dipole-dipole excitation cross section to be
\begin{equation}
 \sigma_{E1E1}= \sigma_0 {N_PZ_P\over A_P} {N_TZ_T\over A_T}
\left({A_P^{1/3}\over A_T^{1/3}} +{A_T^{1/3}\over A_P^{1/3}}\right ),
\end{equation}
with $\sigma_0\approx$ 1.1$\times 10^{-4}$ mb. For $^{12}$C projectiles
on $^{197}$Au, this gives a cross section of .05 mb for simultaneous
excitation, compared to a measured single excitation cross section of
50$\pm$7 mb at 2.1 GeV/nucleon. Numerical integration of the cross section
using the Goldhaber-Teller model\cite{6} confirms the order of magnitude
of this estimate and indicates that the projectile-energy dependence
of the cross section is equally negligible.

	To leading order, then, the simultaneous excitation cross section
is quite negligible, not only because of its $Z_P$ dependence, but also as
a result of the factor of $1/8\pi^3$ coming from phase space. At higher
order in $\alpha$, this situation may change, however, since the
diagram shown in Fig. 2, in which the target and projectile excite
one another via the exchange of two photons, is a classically allowed
process. The resulting cross section will be larger by a factor of
$Z_P^2Z_T^2\alpha^2$ from the leading-order result and may therefore
be significant.

\centerline{\bf {Acknowledgements}}
	This work was performed under the auspices of the U.S. Department of
Energy. One of
us (C.B) would also like to acknowledge support from the Rice University
Physics Department and Bonner Nuclear Laboratory.

\vfill\eject
\centerline{ Figure Captions}
\begin{itemize}
\item{Fig. 1} Feynman diagrams for target excitation (a)with and (b) without
simultaneous excitation of the projectile. A square indicates an elastic form
factor, while circles represent the inelastic transition form factors.

\item{Fig. 2} Feynman diagram for simultaneous excitation of
projectile and target by exchange of two photons.
\end{itemize}
\end{document}